\pdfoutput=1

\documentclass[11pt]{article}

\usepackage{makecell}

\usepackage{xcolor}
\definecolor{darkred}{rgb}{0.8, 0.2, 0.2}
\definecolor{darkgreen}{rgb}{0.2, 0.4, 0.2}

\definecolor{lightblue}{RGB}{221, 231, 245} 
\definecolor{lightyellow}{RGB}{209, 239, 241}
\definecolor{lightgreen}{RGB}{255, 240, 230}
\definecolor{lightred}{RGB}{255,102,102}

\usepackage{pifont}

\usepackage[final]{acl}

\usepackage{times}
\usepackage{latexsym}

\usepackage[T1]{fontenc}

\usepackage[utf8]{inputenc}

\usepackage{microtype}

\usepackage{inconsolata}

\usepackage{graphicx}
\usepackage{booktabs}

\usepackage{subfigure}

\usepackage{tabularx}

\usepackage{xurl}
\usepackage[normalem]{ulem}
\usepackage{amsmath}

\usepackage{multirow}

\usepackage{caption}
\usepackage{soul}
\usepackage{array}
\usepackage{geometry}

\usepackage{tikz}

\usepackage{arydshln}
\usepackage{colortbl}

\usepackage{booktabs}
\usepackage{tabularx}
\usepackage{makecell}
\usepackage{amssymb}    
\usepackage{pifont}      
\usepackage{caption}
\usepackage{graphicx}

\definecolor{lightblue}{RGB}{221, 231, 245} 
\definecolor{lightyellow}{RGB}{209, 239, 241}
\definecolor{lightgreen}{RGB}{255, 240, 230}
\definecolor{lightred}{RGB}{255,102,102}

\title{\raisebox{-0.8em}{\includegraphics[width=3.2em]{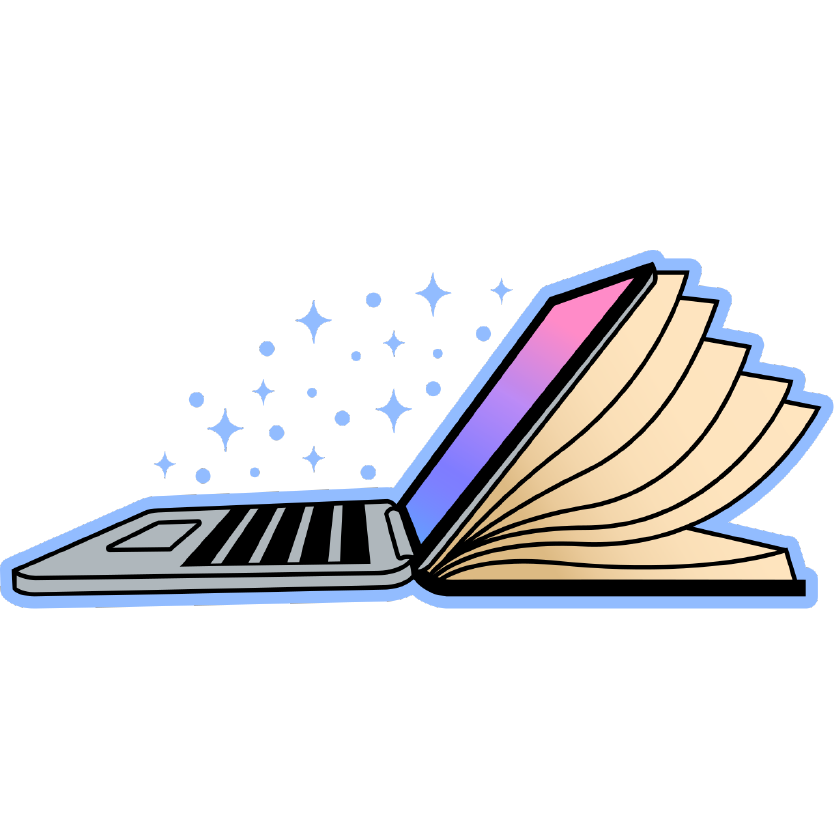}} UltraRAG: A Modular and Automated Toolkit for Adaptive Retrieval-Augmented Generation}

\author{Yuxuan Chen$^{1}$\thanks{\hspace{1mm}Equal Contribution.}, Dewen Guo$^{1}$\footnotemark[1], Sen Mei$^{2}$\footnotemark[1], Xinze Li$^{2}$\footnotemark[1], Hao Chen$^{3}$, Yishan Li$^{1}$, \\ \textbf{Yixuan Wang$^{3}$, Chaoyue Tang$^{1}$, Ruobing Wang$^{4}$, Dingjun Wu$^{1}$, Yukun Yan$^{3}$\thanks{\hspace{1mm}Corresponding Authors.}}\\
\textbf{Zhenghao Liu$^{2}$\footnotemark[2], Shi Yu$^{3}$, Zhiyuan Liu$^{3}$, Maosong Sun$^{3}$}\\
$^1$ModelBest Inc., $^2$Northeastern University, $^3$Tsinghua University\\ $^4$University of Chinese Academy of Sciences\\
}

\begin{document}
\maketitle

\begin{abstract}
Retrieval-Augmented Generation (RAG) significantly enhances the performance of large language models (LLMs) in downstream tasks by integrating external knowledge. To facilitate researchers in deploying RAG systems, various RAG toolkits have been introduced. However, many existing RAG toolkits lack support for knowledge adaptation tailored to specific application scenarios. To address this limitation, we propose UltraRAG, a RAG toolkit that automates knowledge adaptation throughout the entire workflow, from data construction and training to evaluation, while ensuring ease of use. UltraRAG features a user-friendly WebUI that streamlines the RAG process, allowing users to build and optimize systems without coding expertise. It supports multimodal input and provides comprehensive tools for managing the knowledge base. With its highly modular architecture, UltraRAG delivers an end-to-end development solution, enabling seamless knowledge adaptation across diverse user scenarios. The code, demonstration videos, and installable package for UltraRAG are publicly available at \url{https://github.com/OpenBMB/UltraRAG}.
\end{abstract}

\section{Introduction}

Large language models (LLMs)~\cite{achiam2023gpt,touvron2023llama,guo2025deepseek} have demonstrated impressive capabilities in understanding and reasoning. However, due to the limitations of their parameterized knowledge and hallucinations, LLMs usually generate incorrect responses~\cite{guu2020retrieval,ji2023survey,xu2024hallucination}. To address this, retrieval-augmented generation (RAG)~\cite{lewis2020retrieval,guu2020retrieval} has emerged as an effective approach that integrates external knowledge sources, enhancing the accuracy and reliability of responses generated by LLMs. Despite its promising potential, RAG still faces significant challenges in real-world applications. These include the diversity of knowledge corpus formats and modalities~\cite{yu2024visrag}, the complexity of coordinating multiple components~\cite{li2024rag}, and the rapid development of algorithms and models. These challenges create significant obstacles for researchers trying to develop RAG systems.




\newcommand{\yes}{\textcolor{teal}{\textbf{\checkmark}}}
\newcommand{\no}{\textcolor{red!80!black}{\textbf{\texttimes}}}

\begin{table*}[htbp]
\centering
\renewcommand{\arraystretch}{1.4}
\setlength{\tabcolsep}{0.8em}
\small
\begin{tabularx}{\textwidth}{@{} >{\raggedright\arraybackslash}X c c c c c @{}}
\toprule
\textbf{Toolkit} & 
\multicolumn{1}{c}{\textbf{WebUI}} & 
\multicolumn{1}{c}{\textbf{Multimodal}} & 
\multicolumn{1}{c}{\textbf{\makecell{Knowledge\\Management}}} & 
\multicolumn{1}{c}{\textbf{\makecell{End-to-End\\Development}}} & 
\multicolumn{1}{c}{\textbf{\makecell{Knowledge \\Adaptation}}} \\
\midrule
LangChain~\citep{langchain2022}  & \no  & \yes & \yes & \no  & \no  \\
LlamaIndex~\citep{llamaindex}    & \no  & \yes & \yes & \no  & \no  \\
XRAG~\citep{mao2024xrag}     & \yes & \no  & \no  & \no  & \no  \\
FastRAG~\citep{abane2024fastrag} & \no  & \no  & \no  & \no  & \no  \\
RAGLab~\citep{zhang2024raglab}   & \no  & \no  & \yes  & \no  & \no  \\
LocalRQA~\citep{yu2024localrqa}  & \yes  & \no  & \no  & \yes  & \no  \\
FlashRAG~\citep{jin2024flashrag} & \yes & \yes & \no & \yes  & \no  \\
\textbf{UltraRAG (Ours)}         & \yes & \yes & \yes & \yes & \yes \\
\bottomrule
\end{tabularx}

\vspace{-0.5em}  
\caption{Comparison of UltraRAG Features with Other RAG Frameworks.}
\label{tab:rag_comparison}
\end{table*}



For these reasons, a variety of RAG toolkits have been developed to offer technical support for researchers~\cite{llamaindex,langchain2022,jin2024flashrag}. These tools typically modularize the RAG system~\cite{jin2024flashrag}, enabling users to flexibly select and configure different modules, which streamlines both deployment and execution. However, existing RAG toolkits are often overly complex~\cite{langchain2022} and lack knowledge adaptation designs tailored to real-world requirements~\cite{jin2024flashrag}, making it difficult for users to customize and optimize RAG systems for specific scenarios, such as finance and law.

In this paper, we propose UltraRAG, a modular and automated toolkit for adaptive retrieval-augmented generation, enabling users not only to easily deploy and execute RAG systems but also to enhance the RAG pipeline through knowledge adaptation for different scenarios. UltraRAG consists of two global setting modules (Model Management and Knowledge Management) and three core functional modules (Data construction, Training, and Evaluation \& Inference), covering all essential components of the RAG pipeline. From knowledge base preparation to automated data generation, model fine-tuning, and comprehensive evaluation, UltraRAG streamlines the full lifecycle of RAG system development. In summary, UltraRAG provides an end-to-end development platform for RAG systems, facilitating rapid system building, scalable deployment, and fair evaluation. The key features of UltraRAG are outlined in Table~\ref{tab:rag_comparison}, which includes:

\textbf{User-Friendly WebUI.} UltraRAG provides an intuitive WebUI that allows users to easily deploy RAG systems and efficiently process knowledge bases, including encoding and indexing documents in various formats such as TXT, PDF, and Markdown. This user-friendly interface significantly lowers the barrier to usage, allowing individuals with limited technical expertise to quickly build and deploy RAG applications, while reducing both the learning curve and operational complexity.

\textbf{Multimodal.} UltraRAG supports multimodal RAG research and deployment by integrating MLLMs like MiniCPM-V~\cite{yao2024minicpm} and multimodal retrievers~\cite{radford2021learning, zhou2024marvel}. It also incorporates VisRAG~\cite{yu2024visrag}, a model tailored for domain-specific multimodal scenarios, offering comprehensive technical support.

\textbf{Knowledge Management.} UltraRAG enables parameterized knowledge base management, transforming complex processing into simple configurations. Unlike previous methods~\cite{llamaindex,langchain2022} that impose format and specification constraints, UltraRAG supports diverse document formats, simplifying knowledge base processing.

\textbf{End-to-End Development.} UltraRAG offers an end-to-end RAG solution that covers the entire pipeline, from data construction, model fine-tuning to inference and evaluation. It integrates advanced RAG algorithms~\cite{li2024rag,zeng2024kbalign,yu2024visrag}, allowing users to freely combine various techniques and explore numerous configuration possibilities. In addition, UltraRAG includes over 40 widely used datasets, standardized retrieval and generation metrics, and a unified data format. 


\textbf{Knowledge Adaptation.} UltraRAG simplifies the knowledge adaptation process by allowing users to provide only domain-specific corpus. Through its data construction module, the framework automatically generates optimized training data for the entire pipeline, ensuring that the retrieval and generation components are fine-tuned for the specific domain. Our experimental results demonstrate that by adapting knowledge to a particular domain, UltraRAG significantly improves performance, highlighting the advantages of its knowledge adaptation capabilities in real-world applications.

\section{Related Work}

\textbf{Retrieval-Augmented Generation.}
Retrieval-Augmented Generation (RAG) is an effective method for mitigating hallucination and factual inaccuracy issues of large language models (LLMs)~\citep{jiang2023active,xu2023recomp,luo2023sail,hu2023chatdb}.
It has been widely adopted in various natural language processing (NLP) tasks, such as open-domain QA~\citep{trivedi2023interleaving}, language modeling~\citep{he2021efficient}, and dialogue~\citep{cai2019skeleton}. A typical RAG system consists of two key components, a retriever and a generator~\citep{shi2023replug,yu2023augmentation}. The retriever retrieves relevant documents from an external corpus based on the user's query~\citep{DBLP:conf/emnlp/KarpukhinOMLWEC20,DBLP:conf/iclr/XiongXLTLBAO21} and the generator utilizes these documents as context of inputs to augment the generation process~\citep{ram2023context,xu2023recomp}.

With the continuous advancement of research on RAG systems, recent studies have introduced additional modules and explored various training methods specifically tailored for RAG systems~\citep{DBLP:journals/corr/abs-2401-15884,lin2023ra,wang2023learning,wei2024instructrag}. For example, \citet{DBLP:journals/corr/abs-2401-15884} propose an additional retrieval evaluator to refine the quality of retrieved documents. \citet{li2024rag} utilize a rollout method to obtain rewards from the entire RAG system for each module and optimize them based on the reward. The success of these approaches highlights the growing need for a general-purpose RAG toolkit, which can streamline development and evaluation across diverse RAG frameworks.

\noindent \textbf{RAG Toolkits.} 
Various RAG toolkits have been developed to assist users in building customized RAG systems, such as LangChain~\citep{langchain2022} and LlamaIndex~\citep{llamaindex}. These frameworks modularize RAG pipelines and offer seamless integration with knowledge databases, embedding models, and LLM APIs, thereby streamlining development workflows and broadening their range of applications~\citep{haystack}. However, most existing toolkits lack user-friendly WebUIs, do not offer free access to commonly used retrieval corpora, and tend to be overly encapsulated. These limitations significantly hinder their usability and scalability, making them less suitable for both research and practical deployment scenarios~\citep{mao2024xrag,jin2024flashrag}.

To address these limitations, recent work has introduced more transparent and adaptable RAG toolkits. For example, FastRAG~\citep{abane2024fastrag} is built upon Haystack's API, allowing users to freely assemble different modules within RAG pipelines. RAGLAB~\citep{zhang2024raglab} focuses on training RAG systems, offering training strategies tailored for different components. 
However, these toolkits do not adequately support users in end-to-end deployment and development and are not applicable to multimodal tasks. FlashRAG~\citep{jin2024flashrag} not only addresses several of these challenges but also integrates multiple algorithms, allowing users to efficiently reproduce existing methods and explore novel approaches. However, FlashRAG lacks evaluation for different modules in the RAG system~\citep{mao2024xrag} and doesn't support knowledge adaptation for specific scenarios and tasks. In contrast, our proposed UltraRAG toolkit offers an end-to-end modular framework for constructing RAG systems, featuring comprehensive knowledge management and fine-grained evaluation for each module. UltraRAG supports both text and multi-modal tasks, facilitating end-to-end development, evaluation, and deployment of RAG applications.

\begin{figure*}[t]
    \centering
    \includegraphics[width=\linewidth]{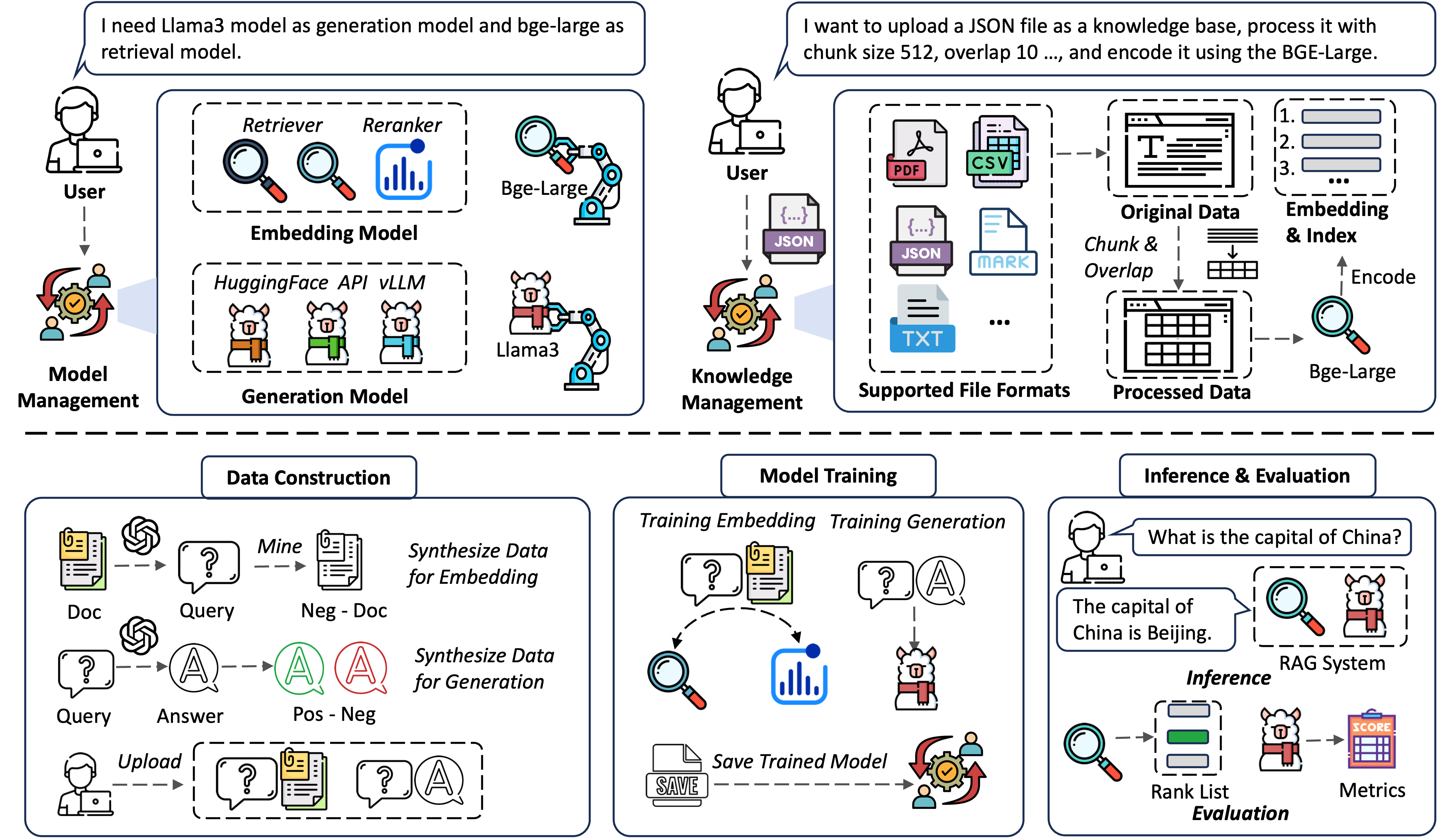}
    \caption{The Overall Architecture of UltraRAG Framework.}
    \label{fig:intro}
\end{figure*}

\section{UltraRAG}

This section introduces UltraRAG, a modular RAG framework designed for the rapid implementation and deployment of RAG pipelines. As shown in Figure~\ref{fig:intro}, UltraRAG consists of two global setting modules: Model Management and Knowledge Management, along with three functional modules: Data Construction, Training, and Evaluation \& Inference. Additionally, UltraRAG offers a user-friendly, visual WebUI, allowing users lacking coding experience to easily process knowledge bases, fine-tune models, and utilize them seamlessly. Details and screenshots of the WebUI can be found in Appendix~\ref{app:screen}.

\subsection{Global Setting}
In this subsection, we introduce the global settings applied in UltraRAG.

\textbf{Model Management.} UltraRAG provides an efficient model management module for the RAG system, enabling the management, deployment, and usage of various models, including retrieval, reranker, and generation models. It supports local models deployed via vLLM~\cite{kwon2023efficient} or HuggingFace Transformers~\cite{wolf2020transformers}, as well as API-based models. To reduce the learning curve, we provide a pre-configured Docker environment and microservices, enabling models to be preloaded in the background for seamless integration with other modules.

\textbf{Knowledge Management.} External knowledge is fundamental to RAG systems, as it provides the documents available for retrieval. However, managing knowledge bases can be challenging, especially for beginners. To address this, UltraRAG offers a user-friendly knowledge management module that enables users to upload knowledge base files in various formats, such as JSON and CSV, and deploy them seamlessly. In addition, users can adjust parameters within the knowledge management module to process the deployed knowledge base, such as adjusting the chunk size, configuring the overlap between chunks, and selecting the embedding model for encoding. Once processed, users can easily obtain the encoded knowledge base along with its corresponding search indexes.

\subsection{Functional Modules}

In this subsection, we introduce several functional modules implemented in UltraRAG, covering the entire workflow of a RAG system.

\textbf{Data Construction.} The data construction module integrates advanced data synthesis techniques~\cite{zhu2024rageval,li2024rag,zeng2024kbalign} to provide datasets to train and evaluate different models within RAG systems. UltraRAG first generates queries automatically based on documents in the user-provided knowledge base. These queries are then used to construct training and evaluation datasets for both retrieval and generation models. For retrieval and reranking models, UltraRAG synthesizes query-document pairs and mines hard negative samples for each query~\cite{xiong2020approximate}. For generation models, it builds on prior work~\cite{li2024rag} to construct supervised fine-tuning (SFT) datasets and direct preference optimization (DPO) datasets, where each query serves as the input for generating high-quality responses. Additionally, users can upload their own pre-constructed datasets and adjust the data proportions to mix different datasets, enabling more effective multi-task training.

\textbf{Training.} Users can further enhance downstream task performance through fine-tuning. Leveraging the training data provided by the data construction module, the training module supports fine-tuning for both embedding models and generation models. Currently, UltraRAG implements two alignment strategies: supervised fine-tuning (SFT) and direct preference optimization (DPO), with plans to incorporate more training strategies in future updates.

\textbf{Evaluation \& Inference.} UltraRAG’s evaluation module provides users with comprehensive methods to assess the performance of both embedding and generation models. It supports a wide range of commonly used retrieval and generation metrics and provides access to over 40 benchmark datasets. In addition, UltraRAG defines a unified data format and allows users to add custom datasets, facilitating flexible and extensible evaluation workflows.

The inference module is available for users who require direct access to RAG services, offering four predefined workflows: Vanilla RAG, KBAlign~\cite{zeng2024kbalign}, VisRAG~\cite{yu2024visrag}, and Adaptive-Note~\cite{wang2024retriever}, while also allowing users to create custom inference workflows according to their specific needs. 
In addition to these workflows, the module supports streaming outputs and visualizes intermediate retrieval and reasoning processes, enabling users to monitor the performance of both retrieval and generation models in real time. Furthermore, UltraRAG supports local deployment via Ollama~\footnote{\url{https://ollama.com/}}, allowing users to deploy models directly on the WebUI as RAG applications that can be fully customized to their specific knowledge bases.

\begin{table}[t]
\centering
\resizebox{\linewidth}{!}{
\begin{tabular}{ll}
\toprule
\textbf{Feature} & \textbf{Model} \\
\midrule
Multimodal & VisRAG \\
Knowledge Management & Adaptive-Note \\
End-to-End Development & RAG-DDR, RA-DIT, KBAlign\\
Knowledge Adaption & RAG-DDR  \\
\bottomrule
\end{tabular}
}
\caption{Typical Implementation. The different features of UltraRAG can support various models and methods.}
\label{tab:tyic_impl}
\end{table}
\subsection{Typical Implementation}

UltraRAG is designed to simplify the research process for researchers by eliminating the need for repetitive development of RAG’s modular components. As shown in Table~\ref{tab:tyic_impl}, we have already implemented several RAG methods, including Vanilla RAG, RA-DIT~\cite{lin2023ra}, Adaptive-Note~\cite{wang2024retriever}, VisRAG~\cite{yu2024visrag}, KBAlign~\cite{zeng2024kbalign}, and RAG-DDR~\cite{li2024rag}. Looking ahead, we plan to expand the framework by incorporating additional RAG baselines. Given the rapid evolution of RAG and the lack of standardized evaluation methods, we believe UltraRAG will accelerate the reproduction of results and enable fair comparisons across different RAG baselines.

\section{Knowledge Adaptation for Legal-Domain RAG System with UltraRAG}

In this section, we present the development of a legal-domain RAG system to highlight UltraRAG’s knowledge adaptation capabilities. We choose the legal domain because the complexity of legal terminology poses significant challenges for LLMs, making domain-specific adaptation essential for improving their real-world applicability.

\subsection{Preliminary of Legal-Domain RAG System}\label{section:4.1}
This subsection introduces the evaluation benchmark, models, and knowledge base used in our experiments. Leveraging UltraRAG’s model and knowledge management modules, we complete the setup effortlessly without coding.

\textbf{Evaluation Dataset.} We select LawBench as the evaluation scenario, a meticulously designed dataset for assessing the legal capabilities of large models within the context of the Chinese legal system. We choose the Scene-based Article Prediction (3-2) and Consultation (3-8) tasks for evaluation. The Scene-based Article Prediction task requires the model to predict relevant legal provisions based on a given scenario and question. The Consultation task focuses on generating appropriate responses to user legal consultations, emphasizing the provision of legal advice based on relevant legal provisions. We choose these tasks because they are more closely aligned with real-world business scenarios.

\textbf{Models.} We use MiniCPM-Embedding-Light\footnote{\url{https://huggingface.co/openbmb/MiniCPM-Embedding-Light}} as the embedding model and MiniCPM-3-4B\footnote{\url{https://huggingface.co/openbmb/MiniCPM3-4B}} as the generation model. These local models are loaded through the model management module to ensure efficient invocation.

\textbf{Knowledge Base.} We have collected a comprehensive knowledge base consisting of over 1,000 law-related books, covering topics such as civil law, criminal law, and judicial cases, to support RAG applications in the legal domain. To optimize retrieval performance and ensure ease of use, we upload all files through the knowledge management module. The index is built by setting the chunk size to 512 with a 15\% overlap, enhancing the retrieval accuracy.

\subsection{Methodology Powered by UltraRAG}\label{section:4.2}

In this subsection, we introduce the methods provided by UltraRAG, which enhance model adaptability and performance in downstream domains through self-constructed training data and model training.



\textbf{Embedding Finetuning.} To finetune the embedding models, we utilize the data construction module of UltraRAG to synthetically generate 2,800 finetuning samples, improving the performance of the MiniCPM-Embedding-Light model. This finetuning process allows the model to better adapt to domain-specific tasks, boosting its retrieval accuracy. The model excels in cross-lingual retrieval, supporting both Chinese and English, and generates high-quality embeddings through bidirectional attention mechanisms and weighted mean pooling. It is also capable of handling long texts of up to 8,192 tokens. To further investigate the training benefits of UltraRAG for the generation models, we use the vanilla MiniCPM-Embedding-Light as the retriever for all the following baselines, and finetune the MiniCPM-3-4B model.

\textbf{Generation Finetuning.} UltraRAG integrates two training methods, UltraRAG-DDR and UltraRAG-KBAlign, to train the generation model.

\textit{UltraRAG-DDR.} We implement UltraRAG-DDR following~\citet{li2024rag}. Leveraging the DDR method provided by UltraRAG’s data construction module, we generate training data by constructing query, ground-truth, and keypoint triplets, and apply data sampling strategies to create diverse responses. Using rule-based rewards, we construct preference pairs, and finally, finetune the model using DPO loss~\cite{rafailov2023direct} with LoRA~\cite{hu2022lora}.

\textit{UltraRAG-KBAlign.} We follow the approach in~\citet{zeng2024kbalign} and apply LoRA-based SFT for finetuning, enhancing the model’s adaptability to the knowledge base through self-supervised learning. The data construction process combines short-range annotations (local information within a single chunk) and long-range annotations (cross-chapter information integration) to improve the model’s knowledge integration capability. Ultimately, during the inference phase, we optimize the aligned knowledge representations to enhance the accuracy and consistency of the generated answers.



\textbf{Inference Workflow.} UltraRAG integrates three different RAG inference workflow.

\textit{UltraRAG-VanillaRAG.} This is a basic RAG workflow, where user task requirements are fulfilled using vanilla retriever and generation model.

\textit{UltraRAG-DeepNote.} We implement the RAG workflow based on~\citet{wang2024retriever}, which uses an adaptive memory reviewer and a stop-exploration strategy to iteratively collect and update knowledge, enhancing both information retrieval and generation quality, and enabling adaptation to more complex user tasks.

\textit{UltraRAG-RAGAdaptation.} In contrast to the previous two workflows, this framework represents the use of models finetuned by the user for knowledge adaptation in specific scenarios. Here, we employ the retrieval and generation models trained with UltraRAG-Embedding and UltraRAG-DDR.

\begin{table}[t]
\centering
\resizebox{\linewidth}{!}{%
\begin{tabular}{lccc}
\hline
Method & MRR@10 & NDCG@10 & Recall@10 \\
\hline
UltraRAG-Embedding & 36.46 & 40.05 & 54.50 \\
w. Finetune & \textbf{37.57} & \textbf{42.12} & \textbf{56.50} \\
\hline
\end{tabular}
}
\caption{Retrieval Performance of UltraRAG with Knowledge Adaptation.}
\label{tab:retrieve}
\end{table}
\subsection{UltraRAG Performance in Legal Scenarios}\label{section:4.3}

In this experiment, we explore the effectiveness of knowledge adaptation in legal scenarios enabled by the UltraRAG training module and assess its effectiveness using the evaluation module. Additionally, we provide case studies in Appendix~\ref{app:case}.

For retrieval evaluation, we use MRR@10, NDCG@10, and Recall@10, assessing MiniCPM-Embedding-Light on 200 GPT-4o-annotated samples before and after finetuning. As shown in Table~\ref{tab:retrieve}, after finetuning on domain-specific data, the performance of the retriever has improved. Through knowledge adaptation in legal scenarios, the retriever better captures the correlation between queries and legal documents, enabling more precise retrieval.

\begin{table}[t]
\centering
\resizebox{\linewidth}{!}{%
\begin{tabular}{lc|lc}
\hline
\multicolumn{2}{c|}{\textbf{Law Prediction (3-2)}} & \multicolumn{2}{c}{\textbf{Consultation (3-8)}} \\
\hline
\textbf{Method} & \textbf{ROUGE-L} & \textbf{Method} & \textbf{ROUGE-L} \\
\hline
VanillaRAG & 40.75 & VanillaRAG & 23.65 \\
KBAlign & 48.72 & DeepNote & 24.62 \\
DDR & \textbf{53.14}& RAGAdaptation & \textbf{25.85} \\
\hline
\end{tabular}
}
\caption{Generation Performance of UltraRAG with Knowledge Adaptation.}
\label{tab:generation}
\end{table}

Next, we explore the performance of the generation model and compare different inference workflows. As shown in Table~\ref{tab:generation}, both KBAlign and DDR demonstrate performance gains, with DDR achieving a 30\% relative improvement. Compared to VanillaRAG, DeepNote significantly enhances performance through its unique memory mechanism, effectively structuring external knowledge. Meanwhile, RAGAdaptation achieves the best results, enabling users to customize and finetune models for specific domains. These findings further highlight the importance of knowledge adaptation and the potential of the UltraRAG framework.

\section{Conclusion}

In this paper, we present UltraRAG, a novel framework designed to address the challenges of domain-specific knowledge adaptation in Retrieval-Augmented Generation systems. With its comprehensive workflow, extensible architecture, and user-friendly WebUI, UltraRAG greatly reduces the barrier to usage, shortens the learning curve, and offers flexible module combinations along with typical implementations, making both the usage and development of RAG systems more adaptable.


\bibliography{acl_latex}

\begin{thebibliography}{44}
\providecommand{\natexlab}[1]{#1}

\bibitem[{Abane et~al.(2024)Abane, Bekri, and Battou}]{abane2024fastrag}
Amar Abane, Anis Bekri, and Abdella Battou. 2024.
\newblock Fastrag: Retrieval augmented generation for semi-structured data.
\newblock \emph{arXiv preprint arXiv:2411.13773}.

\bibitem[{Achiam et~al.(2023)Achiam, Adler, Agarwal, Ahmad, Akkaya, Aleman, Almeida, Altenschmidt, Altman, Anadkat et~al.}]{achiam2023gpt}
Josh Achiam, Steven Adler, Sandhini Agarwal, Lama Ahmad, Ilge Akkaya, Florencia~Leoni Aleman, Diogo Almeida, Janko Altenschmidt, Sam Altman, Shyamal Anadkat, et~al. 2023.
\newblock Gpt-4 technical report.
\newblock \emph{arXiv preprint arXiv:2303.08774}.

\bibitem[{Cai et~al.(2019)Cai, Wang, Bi, Tu, Liu, Lam, and Shi}]{cai2019skeleton}
Deng Cai, Yan Wang, Wei Bi, Zhaopeng Tu, Xiaojiang Liu, Wai Lam, and Shuming Shi. 2019.
\newblock Skeleton-to-response: Dialogue generation guided by retrieval memory.
\newblock In \emph{Proceedings of the 2019 Conference of the North American Chapter of the Association for Computational Linguistics: Human Language Technologies, Volume 1 (Long and Short Papers)}, pages 1219--1228.

\bibitem[{Chase(2022)}]{langchain2022}
Harrison Chase. 2022.
\newblock \href {https://github.com/langchain-ai/langchain} {\href{https://github.com/langchain-ai/langchain}{LangChain}}.

\bibitem[{deepset.ai(2023)}]{haystack}
deepset.ai. 2023.
\newblock \href {https://github.com/deepset-ai/haystack} {\href{https://github.com/deepset-ai/haystack}{Haystack}}.

\bibitem[{Guo et~al.(2025)Guo, Yang, Zhang, Song, Zhang, Xu, Zhu, Ma, Wang, Bi et~al.}]{guo2025deepseek}
Daya Guo, Dejian Yang, Haowei Zhang, Junxiao Song, Ruoyu Zhang, Runxin Xu, Qihao Zhu, Shirong Ma, Peiyi Wang, Xiao Bi, et~al. 2025.
\newblock Deepseek-r1: Incentivizing reasoning capability in llms via reinforcement learning.
\newblock \emph{arXiv preprint arXiv:2501.12948}.

\bibitem[{Guu et~al.(2020)Guu, Lee, Tung, Pasupat, and Chang}]{guu2020retrieval}
Kelvin Guu, Kenton Lee, Zora Tung, Panupong Pasupat, and Mingwei Chang. 2020.
\newblock Retrieval augmented language model pre-training.
\newblock In \emph{International conference on machine learning}, pages 3929--3938. PMLR.

\bibitem[{He et~al.(2021)He, Neubig, and Berg-Kirkpatrick}]{he2021efficient}
Junxian He, Graham Neubig, and Taylor Berg-Kirkpatrick. 2021.
\newblock Efficient nearest neighbor language models.
\newblock In \emph{Proceedings of the 2021 Conference on Empirical Methods in Natural Language Processing}, pages 5703--5714.

\bibitem[{Hu et~al.(2023)Hu, Fu, Du, Luo, Zhao, and Zhao}]{hu2023chatdb}
Chenxu Hu, Jie Fu, Chenzhuang Du, Simian Luo, Junbo Zhao, and Hang Zhao. 2023.
\newblock \href {https://arxiv.org/abs/2306.03901} {Chatdb: Augmenting llms with databases as their symbolic memory}.
\newblock \emph{ArXiv preprint}.

\bibitem[{Hu et~al.(2022)Hu, Shen, Wallis, Allen-Zhu, Li, Wang, Wang, Chen et~al.}]{hu2022lora}
Edward~J Hu, Yelong Shen, Phillip Wallis, Zeyuan Allen-Zhu, Yuanzhi Li, Shean Wang, Lu~Wang, Weizhu Chen, et~al. 2022.
\newblock Lora: Low-rank adaptation of large language models.
\newblock \emph{ICLR}, 1(2):3.

\bibitem[{Ji et~al.(2023)Ji, Lee, Frieske, Yu, Su, Xu, Ishii, Bang, Madotto, and Fung}]{ji2023survey}
Ziwei Ji, Nayeon Lee, Rita Frieske, Tiezheng Yu, Dan Su, Yan Xu, Etsuko Ishii, Ye~Jin Bang, Andrea Madotto, and Pascale Fung. 2023.
\newblock Survey of hallucination in natural language generation.
\newblock \emph{ACM computing surveys}, 55(12):1--38.

\bibitem[{Jiang et~al.(2023)Jiang, Xu, Gao, Sun, Liu, Dwivedi-Yu, Yang, Callan, and Neubig}]{jiang2023active}
Zhengbao Jiang, Frank~F Xu, Luyu Gao, Zhiqing Sun, Qian Liu, Jane Dwivedi-Yu, Yiming Yang, Jamie Callan, and Graham Neubig. 2023.
\newblock \href {https://aclanthology.org/2023.emnlp-main.495.pdf} {Active retrieval augmented generation}.
\newblock In \emph{Proceedings of EMNLP}, pages 7969--7992.

\bibitem[{Jin et~al.(2024)Jin, Zhu, Yang, Zhang, and Dou}]{jin2024flashrag}
Jiajie Jin, Yutao Zhu, Xinyu Yang, Chenghao Zhang, and Zhicheng Dou. 2024.
\newblock Flashrag: A modular toolkit for efficient retrieval-augmented generation research.
\newblock \emph{arXiv preprint arXiv:2405.13576}.

\bibitem[{Karpukhin et~al.(2020)Karpukhin, Oguz, Min, Lewis, Wu, Edunov, Chen, and Yih}]{DBLP:conf/emnlp/KarpukhinOMLWEC20}
Vladimir Karpukhin, Barlas Oguz, Sewon Min, Patrick Lewis, Ledell Wu, Sergey Edunov, Danqi Chen, and Wen-tau Yih. 2020.
\newblock \href {https://aclanthology.org/2020.emnlp-main.550} {Dense passage retrieval for open-domain question answering}.
\newblock In \emph{Proceedings of EMNLP}, pages 6769--6781.

\bibitem[{Kwon et~al.(2023)Kwon, Li, Zhuang, Sheng, Zheng, Yu, Gonzalez, Zhang, and Stoica}]{kwon2023efficient}
Woosuk Kwon, Zhuohan Li, Siyuan Zhuang, Ying Sheng, Lianmin Zheng, Cody~Hao Yu, Joseph Gonzalez, Hao Zhang, and Ion Stoica. 2023.
\newblock Efficient memory management for large language model serving with pagedattention.
\newblock In \emph{Proceedings of the 29th Symposium on Operating Systems Principles}, pages 611--626.

\bibitem[{Lewis et~al.(2020)Lewis, Perez, Piktus, Petroni, Karpukhin, Goyal, K{\"u}ttler, Lewis, Yih, Rockt{\"a}schel et~al.}]{lewis2020retrieval}
Patrick Lewis, Ethan Perez, Aleksandra Piktus, Fabio Petroni, Vladimir Karpukhin, Naman Goyal, Heinrich K{\"u}ttler, Mike Lewis, Wen-tau Yih, Tim Rockt{\"a}schel, et~al. 2020.
\newblock Retrieval-augmented generation for knowledge-intensive nlp tasks.
\newblock \emph{Advances in neural information processing systems}, 33:9459--9474.

\bibitem[{Li et~al.(2024)Li, Mei, Liu, Yan, Wang, Yu, Zeng, Chen, Yu, Liu et~al.}]{li2024rag}
Xinze Li, Sen Mei, Zhenghao Liu, Yukun Yan, Shuo Wang, Shi Yu, Zheni Zeng, Hao Chen, Ge~Yu, Zhiyuan Liu, et~al. 2024.
\newblock Rag-ddr: Optimizing retrieval-augmented generation using differentiable data rewards.
\newblock \emph{arXiv preprint arXiv:2410.13509}.

\bibitem[{Lin et~al.(2023)Lin, Chen, Chen, Shi, Lomeli, James, Rodriguez, Kahn, Szilvasy, Lewis et~al.}]{lin2023ra}
Xi~Victoria Lin, Xilun Chen, Mingda Chen, Weijia Shi, Maria Lomeli, Rich James, Pedro Rodriguez, Jacob Kahn, Gergely Szilvasy, Mike Lewis, et~al. 2023.
\newblock \href {https://arxiv.org/abs/2310.01352} {Ra-dit: Retrieval-augmented dual instruction tuning}.
\newblock \emph{ArXiv preprint}.

\bibitem[{Liu(2022)}]{llamaindex}
Jerry Liu. 2022.
\newblock \href {https://github.com/run-llama/llama_index} {\href{https://github.com/run-llama/llama_index}{LlamaIndex}}.

\bibitem[{Luo et~al.(2023)Luo, Chuang, Gong, Zhang, Kim, Wu, Fox, Meng, and Glass}]{luo2023sail}
Hongyin Luo, Yung-Sung Chuang, Yuan Gong, Tianhua Zhang, Yoon Kim, Xixin Wu, Danny Fox, Helen Meng, and James Glass. 2023.
\newblock \href {https://arxiv.org/abs/2305.15225} {Sail: Search-augmented instruction learning}.
\newblock \emph{ArXiv preprint}.

\bibitem[{Mao et~al.(2024)Mao, Luo, Zhang, Hao, Cao, Wang, Guan, Huang, Jiang, Guo et~al.}]{mao2024xrag}
Qianren Mao, Yangyifei Luo, Jinlong Zhang, Hanwen Hao, Zhilong Cao, Xiaolong Wang, Xiao Guan, Zhenting Huang, Weifeng Jiang, Shuyu Guo, et~al. 2024.
\newblock Xrag: examining the core--benchmarking foundational components in advanced retrieval-augmented generation.
\newblock \emph{arXiv preprint arXiv:2412.15529}.

\bibitem[{Radford et~al.(2021)Radford, Kim, Hallacy, Ramesh, Goh, Agarwal, Sastry, Askell, Mishkin, Clark et~al.}]{radford2021learning}
Alec Radford, Jong~Wook Kim, Chris Hallacy, Aditya Ramesh, Gabriel Goh, Sandhini Agarwal, Girish Sastry, Amanda Askell, Pamela Mishkin, Jack Clark, et~al. 2021.
\newblock Learning transferable visual models from natural language supervision.
\newblock In \emph{International conference on machine learning}, pages 8748--8763. PmLR.

\bibitem[{Rafailov et~al.(2023)Rafailov, Sharma, Mitchell, Manning, Ermon, and Finn}]{rafailov2023direct}
Rafael Rafailov, Archit Sharma, Eric Mitchell, Christopher~D Manning, Stefano Ermon, and Chelsea Finn. 2023.
\newblock Direct preference optimization: Your language model is secretly a reward model.
\newblock \emph{Advances in Neural Information Processing Systems}, 36:53728--53741.

\bibitem[{Ram et~al.(2023)Ram, Levine, Dalmedigos, Muhlgay, Shashua, Leyton-Brown, and Shoham}]{ram2023context}
Ori Ram, Yoav Levine, Itay Dalmedigos, Dor Muhlgay, Amnon Shashua, Kevin Leyton-Brown, and Yoav Shoham. 2023.
\newblock \href {https://arxiv.org/abs/2302.00083} {In-context retrieval-augmented language models}.
\newblock \emph{Transactions of the Association for Computational Linguistics}, pages 1316--1331.

\bibitem[{Shi et~al.(2023)Shi, Min, Yasunaga, Seo, James, Lewis, Zettlemoyer, and Yih}]{shi2023replug}
Weijia Shi, Sewon Min, Michihiro Yasunaga, Minjoon Seo, Rich James, Mike Lewis, Luke Zettlemoyer, and Wen-tau Yih. 2023.
\newblock \href {https://arxiv.org/abs/2301.12652} {Replug: Retrieval-augmented black-box language models}.
\newblock \emph{ArXiv preprint}.

\bibitem[{Touvron et~al.(2023)Touvron, Lavril, Izacard, Martinet, Lachaux, Lacroix, Rozi{\`e}re, Goyal, Hambro, Azhar et~al.}]{touvron2023llama}
Hugo Touvron, Thibaut Lavril, Gautier Izacard, Xavier Martinet, Marie-Anne Lachaux, Timoth{\'e}e Lacroix, Baptiste Rozi{\`e}re, Naman Goyal, Eric Hambro, Faisal Azhar, et~al. 2023.
\newblock Llama: Open and efficient foundation language models.
\newblock \emph{arXiv preprint arXiv:2302.13971}.

\bibitem[{Trivedi et~al.(2023)Trivedi, Balasubramanian, Khot, and Sabharwal}]{trivedi2023interleaving}
Harsh Trivedi, Niranjan Balasubramanian, Tushar Khot, and Ashish Sabharwal. 2023.
\newblock Interleaving retrieval with chain-of-thought reasoning for knowledge-intensive multi-step questions.
\newblock In \emph{Proceedings of the 61st Annual Meeting of the Association for Computational Linguistics (Volume 1: Long Papers)}, pages 10014--10037.

\bibitem[{Wang et~al.(2024)Wang, Zha, Yu, Zhao, Chen, Wang, Wang, Yan, Liu, Han et~al.}]{wang2024retriever}
Ruobing Wang, Daren Zha, Shi Yu, Qingfei Zhao, Yuxuan Chen, Yixuan Wang, Shuo Wang, Yukun Yan, Zhenghao Liu, Xu~Han, et~al. 2024.
\newblock Retriever-and-memory: Towards adaptive note-enhanced retrieval-augmented generation.
\newblock \emph{arXiv preprint arXiv:2410.08821}.

\bibitem[{Wang et~al.(2023)Wang, Araki, Jiang, Parvez, and Neubig}]{wang2023learning}
Zhiruo Wang, Jun Araki, Zhengbao Jiang, Md~Rizwan Parvez, and Graham Neubig. 2023.
\newblock Learning to filter context for retrieval-augmented generation.
\newblock \emph{arXiv preprint arXiv:2311.08377}.

\bibitem[{Wei et~al.(2024)Wei, Chen, and Meng}]{wei2024instructrag}
Zhepei Wei, Wei-Lin Chen, and Yu~Meng. 2024.
\newblock Instructrag: Instructing retrieval-augmented generation with explicit denoising.
\newblock \emph{arXiv e-prints}, pages arXiv--2406.

\bibitem[{Wolf et~al.(2020)Wolf, Debut, Sanh, Chaumond, Delangue, Moi, Cistac, Rault, Louf, Funtowicz et~al.}]{wolf2020transformers}
Thomas Wolf, Lysandre Debut, Victor Sanh, Julien Chaumond, Clement Delangue, Anthony Moi, Pierric Cistac, Tim Rault, R{\'e}mi Louf, Morgan Funtowicz, et~al. 2020.
\newblock Transformers: State-of-the-art natural language processing.
\newblock In \emph{Proceedings of the 2020 conference on empirical methods in natural language processing: system demonstrations}, pages 38--45.

\bibitem[{Xiong et~al.(2020)Xiong, Xiong, Li, Tang, Liu, Bennett, Ahmed, and Overwijk}]{xiong2020approximate}
Lee Xiong, Chenyan Xiong, Ye~Li, Kwok-Fung Tang, Jialin Liu, Paul Bennett, Junaid Ahmed, and Arnold Overwijk. 2020.
\newblock Approximate nearest neighbor negative contrastive learning for dense text retrieval.
\newblock \emph{arXiv preprint arXiv:2007.00808}.

\bibitem[{Xiong et~al.(2021)Xiong, Xiong, Li, Tang, Liu, Bennett, Ahmed, and Overwijk}]{DBLP:conf/iclr/XiongXLTLBAO21}
Lee Xiong, Chenyan Xiong, Ye~Li, Kwok{-}Fung Tang, Jialin Liu, Paul~N. Bennett, Junaid Ahmed, and Arnold Overwijk. 2021.
\newblock \href {https://arxiv.org/pdf/2007.00808} {Approximate nearest neighbor negative contrastive learning for dense text retrieval}.
\newblock In \emph{Proceedings of ICLR}.

\bibitem[{Xu et~al.(2023)Xu, Shi, and Choi}]{xu2023recomp}
Fangyuan Xu, Weijia Shi, and Eunsol Choi. 2023.
\newblock \href {https://arxiv.org/abs/2310.04408} {Recomp: Improving retrieval-augmented lms with compression and selective augmentation}.
\newblock \emph{ArXiv preprint}.

\bibitem[{Xu et~al.(2024)Xu, Jain, and Kankanhalli}]{xu2024hallucination}
Ziwei Xu, Sanjay Jain, and Mohan Kankanhalli. 2024.
\newblock Hallucination is inevitable: An innate limitation of large language models.
\newblock \emph{arXiv preprint arXiv:2401.11817}.

\bibitem[{Yan et~al.(2024)Yan, Gu, Zhu, and Ling}]{DBLP:journals/corr/abs-2401-15884}
Shi{-}Qi Yan, Jia{-}Chen Gu, Yun Zhu, and Zhen{-}Hua Ling. 2024.
\newblock \href {https://arxiv.org/abs/2401.15884} {Corrective retrieval augmented generation}.

\bibitem[{Yao et~al.(2024)Yao, Yu, Zhang, Wang, Cui, Zhu, Cai, Li, Zhao, He et~al.}]{yao2024minicpm}
Yuan Yao, Tianyu Yu, Ao~Zhang, Chongyi Wang, Junbo Cui, Hongji Zhu, Tianchi Cai, Haoyu Li, Weilin Zhao, Zhihui He, et~al. 2024.
\newblock Minicpm-v: A gpt-4v level mllm on your phone.
\newblock \emph{arXiv preprint arXiv:2408.01800}.

\bibitem[{Yu et~al.(2024{\natexlab{a}})Yu, Tang, Xu, Cui, Ran, Yan, Liu, Wang, Han, Liu et~al.}]{yu2024visrag}
Shi Yu, Chaoyue Tang, Bokai Xu, Junbo Cui, Junhao Ran, Yukun Yan, Zhenghao Liu, Shuo Wang, Xu~Han, Zhiyuan Liu, et~al. 2024{\natexlab{a}}.
\newblock Visrag: Vision-based retrieval-augmented generation on multi-modality documents.
\newblock \emph{arXiv preprint arXiv:2410.10594}.

\bibitem[{Yu et~al.(2024{\natexlab{b}})Yu, Lu, and Yu}]{yu2024localrqa}
Xiao Yu, Yunan Lu, and Zhou Yu. 2024{\natexlab{b}}.
\newblock Localrqa: From generating data to locally training, testing, and deploying retrieval-augmented qa systems.
\newblock \emph{arXiv preprint arXiv:2403.00982}.

\bibitem[{Yu et~al.(2023)Yu, Xiong, Yu, and Liu}]{yu2023augmentation}
Zichun Yu, Chenyan Xiong, Shi Yu, and Zhiyuan Liu. 2023.
\newblock \href {https://arxiv.org/abs/2305.17331} {Augmentation-adapted retriever improves generalization of language models as generic plug-in}.
\newblock \emph{ArXiv preprint}.

\bibitem[{Zeng et~al.(2024)Zeng, Chen, Yu, Yan, Liu, Wang, Han, Liu, and Sun}]{zeng2024kbalign}
Zheni Zeng, Yuxuan Chen, Shi Yu, Yukun Yan, Zhenghao Liu, Shuo Wang, Xu~Han, Zhiyuan Liu, and Maosong Sun. 2024.
\newblock Kbalign: Kbalign: Efficient self adaptation on specific knowledge bases.
\newblock \emph{arXiv preprint arXiv:2411.14790}.

\bibitem[{Zhang et~al.(2024)Zhang, Song, Wang, Tang, Li, Zeng, Wu, Ye, Xu, Zhang et~al.}]{zhang2024raglab}
Xuanwang Zhang, Yunze Song, Yidong Wang, Shuyun Tang, Xinfeng Li, Zhengran Zeng, Zhen Wu, Wei Ye, Wenyuan Xu, Yue Zhang, et~al. 2024.
\newblock Raglab: A modular and research-oriented unified framework for retrieval-augmented generation.
\newblock \emph{arXiv preprint arXiv:2408.11381}.

\bibitem[{Zhou et~al.(2024)Zhou, Mei, Li, Liu, Xiong, Liu, Gu, and Yu}]{zhou2024marvel}
Tianshuo Zhou, Sen Mei, Xinze Li, Zhenghao Liu, Chenyan Xiong, Zhiyuan Liu, Yu~Gu, and Ge~Yu. 2024.
\newblock Marvel: Unlocking the multi-modal capability of dense retrieval via visual module plugin.
\newblock In \emph{Proceedings of the 62nd Annual Meeting of the Association for Computational Linguistics (Volume 1: Long Papers)}, pages 14608--14624.

\bibitem[{Zhu et~al.(2024)Zhu, Luo, Xu, Wang, Yu, Wang, Yan, Liu, Han, Liu et~al.}]{zhu2024rageval}
Kunlun Zhu, Yifan Luo, Dingling Xu, Ruobing Wang, Shi Yu, Shuo Wang, Yukun Yan, Zhenghao Liu, Xu~Han, Zhiyuan Liu, et~al. 2024.
\newblock Rageval: Scenario specific rag evaluation dataset generation framework.
\newblock \emph{arXiv preprint arXiv:2408.01262}.

\end{thebibliography}

\clearpage
\newpage
\appendix
\section{Appendix}

\begin{table}[t]
\small
\renewcommand{\arraystretch}{1.2}

\definecolor{titlecolor}{RGB}{248, 249, 253}

\setlength{\arrayrulewidth}{0.7pt}

\begin{tabular}{|p{0.95\columnwidth}|}
\hline  
\cellcolor{titlecolor}\textbf{Case \#1} \\ \hline
\textbf{Instruction:} Please provide the legal basis according to the specific scenario and issue, only including the relevant article of law, with each scenario involving only one article.\\
\textbf{Query:} Scenario: Xiao Ming works at a foreign-invested enterprise and suffers a work injury due to an unsafe working environment. According to which law should the foreign-invested enterprise comply with labor protection regulations?\\
\textbf{Ground Truth:} \sethlcolor{lightblue}\hl{According to Article 32 of the Foreign Investment Law}, foreign-invested enterprises shall comply with the relevant labor protection provisions of laws and administrative regulations when carrying out production and business activities. \sethlcolor{lightblue}\hl{Therefore, foreign-invested enterprises should ensure a safe working environment for employees and comply with national labor protection laws and regulations.} When employees are injured due to work-related reasons, the enterprise should provide appropriate medical treatment and actively assist employees in applying for work-related injury insurance benefits.\\
\textbf{VanillaRAG Predit:} \sethlcolor{lightgreen}\hl{According to Article 38 of the Labor Law of the People’s Republic of China}, workers have the right to social insurance and benefits. Workers are entitled to medical leave if they are ill or injured not due to work, and during the treatment for work-related injuries, their original salary should remain unchanged. The employer should pay the relevant benefits based on the principle of work-related injury.\\
\textbf{UltraRAG-RAGAdaption Predit:} \sethlcolor{lightyellow}\hl{According to Article 32 of the Regulations on the Administration of Foreign-invested Enterprises}, foreign-invested enterprises must comply with the relevant laws and administrative regulations concerning labor protection and social insurance. \sethlcolor{lightyellow}\hl{Therefore, foreign-invested enterprises should adhere to labor protection regulations, ensure a safe working environment, and prevent work-related accidents.}\\
\hline
\cellcolor{titlecolor}\textbf{Case \#2} \\ \hline
\textbf{Instruction:} Please provide the specific scenario and question, and I will give you the relevant legal provision. Each scenario will be linked to a single legal article.\\
\textbf{Query:} In an auction, a bidder bids a high price for a painting, and the auctioneer confirms the highest bid by dropping the hammer. According to the law, what is the status of the auction once the bidder’s highest offer is confirmed?\\
\textbf{Ground Truth:} \sethlcolor{lightblue}\hl{According to Article 51 of the Auction Law}, once the highest bid of the bidder is confirmed by the auctioneer’s hammer or other public methods of indicating acceptance, \sethlcolor{lightblue}\hl{the auction is concluded}. Therefore, in this scenario, after the auctioneer’s confirmation, the bidder’s highest offer is finalized and publicly announced, and the auction is deemed to have been successfully concluded.\\
\textbf{VanillaRAG Predit:} \sethlcolor{lightgreen}\hl{The auction is concluded.}\\
\textbf{UltraRAG-RAGAdaption Predit:} \sethlcolor{lightyellow}\hl{According to Article 51 of the Auction Law}, once the highest bid of the bidder is confirmed by the auctioneer’s hammer or other public means of indicating that the sale is final, \sethlcolor{lightyellow}\hl{the status of the auction is ``auction concluded''}.\\
\hline
\end{tabular}
\caption{Case Studies. Different colors are used to annotate these matched phrases for each model: \sethlcolor{lightblue}\hl{Blue} for Ground Truth, \sethlcolor{lightgreen}\hl{Orange} for Vanilla RAG and \sethlcolor{lightyellow}\hl{Green} for RAGAdaption.}
\label{tab:case_study}
\end{table}
\subsection{Case Study}\label{app:case}
In Table~\ref{tab:case_study}, we present two cases to further demonstrate the effectiveness of UltraRAG’s knowledge adaptation.

As shown in the first case, which addresses labor protection issues in foreign-invested enterprises, VanillaRAG mistakenly references Article 38 of the Labor Law. However, the core issue relates to the responsibilities of foreign-invested enterprises, not general labor law. After being fine-tuned for legal domain-specific tasks, UltraRAG accurately understands the issue and correctly cites Article 32 of the Administrative Regulations on Foreign-invested Enterprises, providing the appropriate response.

In the second case, both models offer a correct response. However, VanillaRAG simply states ``auction concluded'' without referencing the specific legal provision. In contrast, UltraRAG delivers a more accurate response by citing the relevant legal article, better addressing the user’s needs.

\begin{figure*}
\centering
\subfigure[Global Setting.]{
\label{fig:screenshot:glo}
\begin{minipage}[b]{1\textwidth}
\centering
\includegraphics[width=1\textwidth]{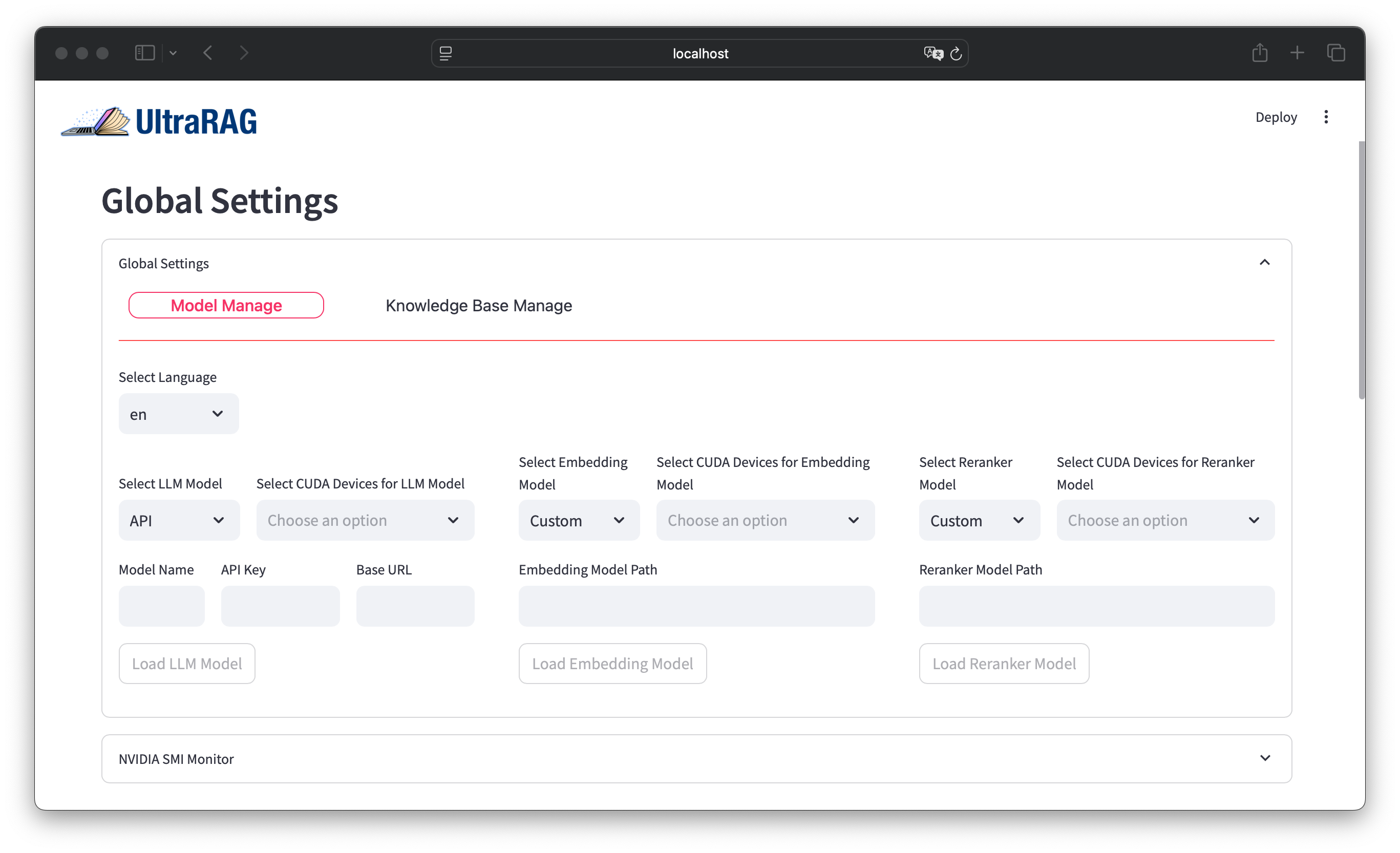}
\end{minipage}
}
\subfigure[UltraRAG DeepNote.]{
\label{fig:screenshot:deep}
\begin{minipage}[b]{1\textwidth}
\centering
\includegraphics[width=0.97\textwidth]{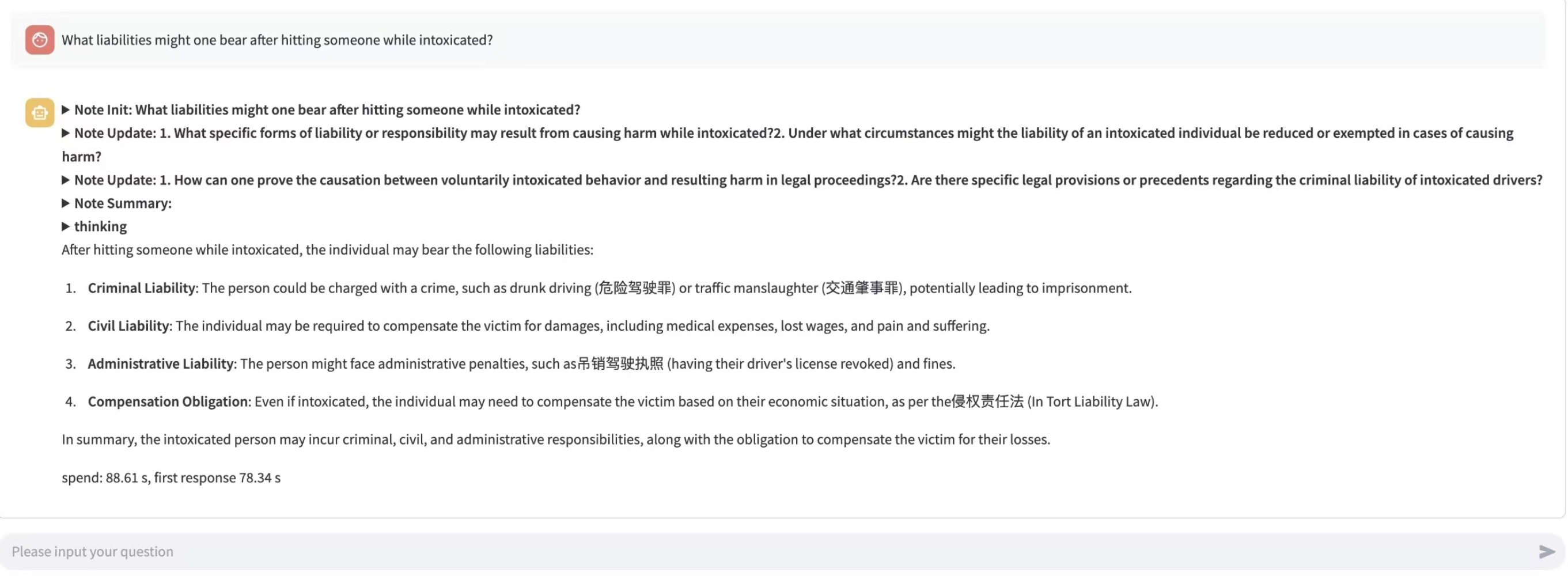}
\end{minipage}
}
\caption{Screenshots of UltraRAG.} \label{fig:screenshot}
\end{figure*}
\subsection{Screenshots of UltraRAG}\label{app:screen}

In this subsection, we begin by introducing the WebUI interface of UltraRAG, highlighting its key features and functionalities, followed by a detailed description of the DeepNote workflow.

As shown in Figure~\ref{fig:screenshot:glo}, the global settings interface enables seamless navigation between the model management and knowledge management modules. Within the model management module, users can easily select the desired model, specify the model path, assign a CUDA device, and configure other essential settings. These features streamline the deployment process, allowing users to quickly set up a RAG system.

Next, we describe the workflow of UltraRAG DeepNote. As illustrated in Figure~\ref{fig:screenshot:deep}, UltraRAG DeepNote introduces an advanced memory mechanism that dynamically organizes and updates the retrieved knowledge. Unlike VanillaRAG, which directly concatenates retrieved documents to the context, UltraRAG DeepNote continuously organizes knowledge throughout the inference process. This approach enhances knowledge structuring and improves information integration, leading to more accurate and coherent responses.


\end{document}